\documentclass{article}

\usepackage{arxiv}

\usepackage[utf8]{inputenc} 
\usepackage[T1]{fontenc}    
\usepackage{hyperref}       
\usepackage{url}            
\usepackage{booktabs}       
\usepackage{amsfonts}       
\usepackage{nicefrac}       
\usepackage{microtype}      
\usepackage{lipsum}
\usepackage{graphicx}
\graphicspath{ {./images/} }

\title{CyMed: A Framework for Testing Cybersecurity of Connected Medical Devices}

\author{
 Christopher Scherb \\
  University of Applied Sciences and Arts,\\
  Northwestern Switzerland\\
  \texttt{christopher.scherb@fhnw.ch} \\
   \And
 Adrian Hadayah \\
University of Applied Sciences and Arts, \\
Northwestern Switzerland \\
  \texttt{ahadayah@sap-all.com} \\
  \And
 Luc Bryan Heitz \\
University of Applied Sciences and Arts, \\
Northwestern Switzerland \\
  \texttt{luc.heitz@fhnw.ch} \\
}

\begin{document}
\maketitle
\begin{abstract}
Connected Medical Devices (CMDs) have a large impact on patients as they allow them to lead a more normal life. Any malfunction could not only remove the health benefits the CMDs provide, they could also cause further harm to the patient.  Due to this, there are many safety regulations which must be adhered to prior to a CMD entering the market. 
However, while many detailed safety regulations exist, there are a fundamental lack of cybersecurity frameworks applicable to CMDs.  While there are recent regulations which aim to enforce cybersecurity practices, they are vague and do not contain the concrete steps necessary to implement cybersecurity.
This paper aims to fill that gap by describing a framework, CyMed, to be used by vendors and ens-users, which contains concrete measures to improve the resilience of CMDs against cyber attack.  The CyMed framework is subsequently evaluated based on practical tests as well as expert interviews.
\end{abstract}


\section{Introduction}
Protecting our increasingly connected infrastructure is one of the core challenges of our time. Especially critical infrastructure such as power plants, industrial production and medical devices are getting more and more in the focus of attackers. 
To protect CMDs different approaches are chosen. Ideally, if any vulnerability gets public, the devices should be patched as fast as possible. 
However, due to the strict safety regulations and certifications a CMD cannot be patched as fast as a normal computer since the patch needs to pass all certifications. Therefore, different approaches are required to protect CMDs as well as to increase the cybersecurity of critical infrastructure in general. To reduce the attack surface against running CMDs, network segmentation is one of the commonly used tools, where the access to static CMDs (such as MRI, CT, X-ray) is limited to devices that require the communication. However, network segmentation as well as any form of network surveillance are not a cure but rather fighting symptoms of insecure devices. A more sustainable approach to increase the cybersecurity of CMDs is to ensure they are designed to be secure and contain as little vulnerabilities as possible. The EMA and the FDA issue guidelines and regulations for the development of medical devices and CMDs such as the general regulation for medical devices (EU: EU 2017/745 and EU 2017/746, FDA: 21 CFR 820 and 21 CFR 11) as well as the guidelines for cybersecurity (MDCG 2019-16 and \emph{Content of Premarket Submissions for Management of Cybersecurity in Medical Devices (FDA)}). While these guidelines for cybersecurity are a good starting point, their main content is about risk management and capabilities which should be fulfilled by the CMDs, for example \emph{Automatic Logoff}, \emph{Person Authentication} and \emph{System and OS Hardening}. However, it is completely missing how especially technical measures can concretely be implemented and tested by the developers. In our point of view, this is a major flaw, since many companies developing CMDs are lacking resources and knowledge how to secure devices. 
Moreover, IEC 62304 which defines the software life cycle process for medical devices does not cover cybersecurity yet.  
Therefore, in this paper we propose a framework called CyMed with concrete measure that can be applied to CMDs to increase their cybersecurity and remove bugs and vulnerabilities before the devices enter the market. CyMed focuses on Software and Medical devices in two ways: 
First on \emph{Software in a Medical Device} and second on \emph{Software as a Medical Device (SaMD)}. Software in a Medical Device refers to any software which is required to run a CMD such as the software in a motion tracker, a pacemaker, an insulin pump or an MRI~(\cite{gordon2019challenges, scherb2018resolution}). SaMD refers to software such as mobile apps or computer programs which perform the purposes of a medical device without being part of the hardware of a medical device, for example a fitness tracking app, the software on a computer connecting to an MRI to show the scans, the software used to calibrate a pacemaker or also connector software which connects physicians to insurance companies etc~(\cite{carroll2016software}). 
We evaluate our CyMed-framework both with practical tests as well as with expert interviews.
The rest of the paper is structured as follows: In section \ref{sec:ProblemStatement} we summarize our problem statement, in section \ref{sec:backrelated} we present related work and technologies as well as further background information, followed by the methodology in section \ref{sec:methodology}.
In section \ref{sec:framework} we present our CyMed-framework for testing CMDs and afterwards we present our evaluation (section \ref{sec:eval}), before we conclude our paper. 

\section{Problem Statement}
\label{sec:ProblemStatement}
Cybersecurity for CMDs is a critical topic, during the development phase and in operation. Generally, CMDs are strongly regulated devices, as described in section \ref{sec:backrelated}, and there are a lot of guidelines and standards. Naturally, guidelines and standards are not designed to give detailed information about how a state-of-the-art security check is done but more abstract and guiding information which are more timeless. However, for manufacturers it is important to assess the state of the art possibilities for increasing the cybersecurity of their CMDs. We fill the gap by providing a framework giving tips and methods how to improve security of CMDs and how to protect them from cyber attacks. 
In this paper we answer the research questions:
\begin{itemize}
    \item \emph{How can the resistance of CMDs against cyber attacks be increased?}
    \item \emph{Which concrete measure can be taken during and after the development process to improve the resistance against cyber attacks?}
\end{itemize}

\section{Background \& Related Work}
\label{sec:backrelated}
In the following we will present details about existing guidance for cybersecurity of medical devices as well as methods we apply in our CyMed-framework. 

\subsection{Regulatory Guidance for Cybersecurity of Connected Medical Devices}
\begin{figure}
    \centering
    \includegraphics[width=0.38\textwidth]{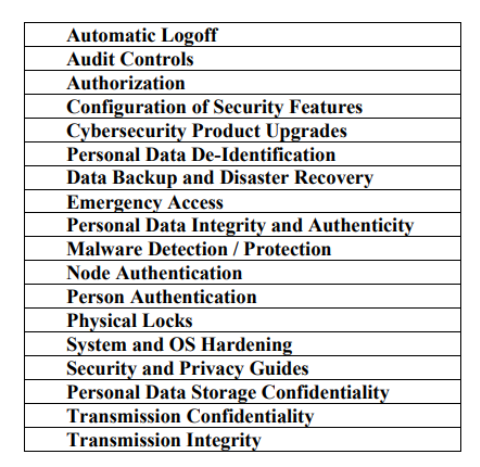}
    \caption{Indicative List of security capabilities for CMDs}
    \label{fig:mdcg_cap}
\end{figure}

MDCG 2019-16\footnote{\url{https://health.ec.europa.eu/system/files/2022-01/md_cybersecurity_en.pdf}}, Guidance on Cybersecurity for medical devices is a guidance developed by the Medical Device Coordination Group (MDCG) of the European Union, consisting of experts from all member states~(\cite{granlund2021medical}). The guidance is one of the most detailed documents about cybersecurity for CMDs and comes up with many recommendations ranging from basic cybersecurity up to guides for secure design and manufacture and post market surveillance. The responsibility of keeping the medical devices secure is shared between the different stakeholders, mainly the operator and the developer. 
The guidance recommends manufactures of CMDs to consider IT-security, operation security as well as information security. IT-Security is defined as the general protection from disruption or misdirection of the services. Operation security is the protection against intended corruption of procedures or workflows and information security is the "Protection against the threat of theft, deletion or alteration of stored or transmitted data within a cyber system".
The risk of the devices needs to be reduced to a reasonable/foreseeable risk in all operation modes. Therefore, secure design and manufacturing should be established, by having risk management processes through the entire life cycle of the device, and by protecting the CMD against unauthorized access. Furthermore, threats and possible vulnerabilities should be identified and risk control measures should be taken. An important issue that needs to be addressed is the relation of cybersecurity and safety. Too strict or too loose cybersecurity may have an impact on the safety of the CMD, by either blocking functions required for the safe operation or allowing attackers to manipulate the device. 
MDCG 2019-16 devices capabilities that should be fulfilled by a medical device to be secure as shown in figure \ref{fig:mdcg_cap} as well as examples for cyber risks. 

\begin{figure}
    \centering
    \includegraphics[width=0.48\textwidth]{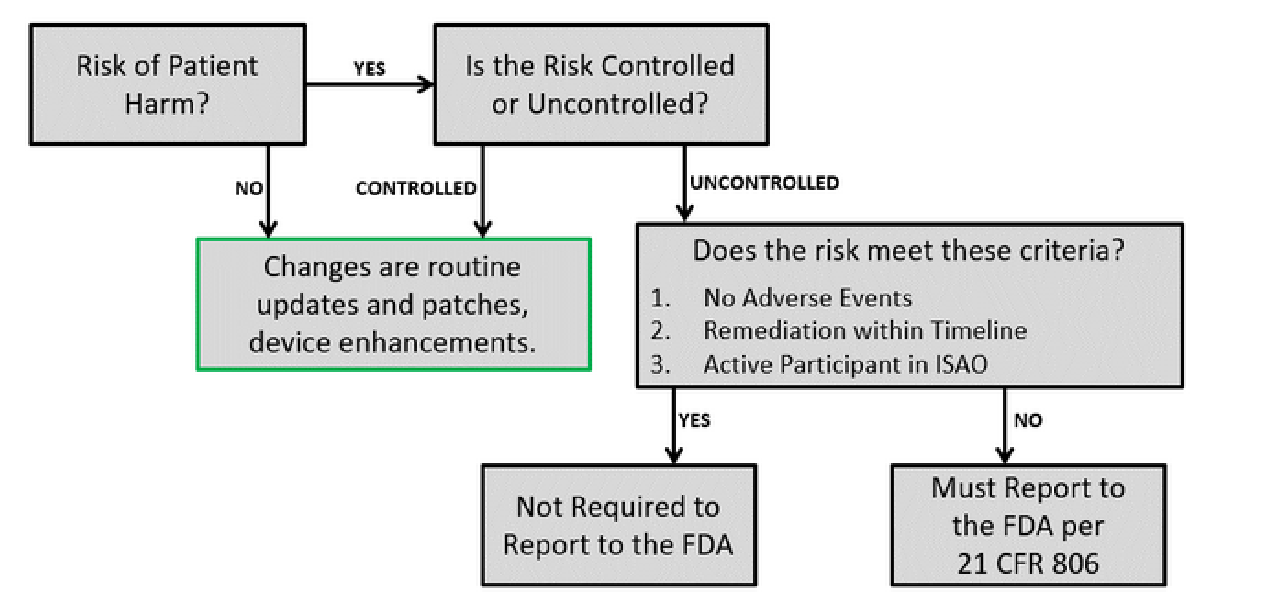}
    \caption{FDA risk assessment}
    \label{fig:fdarisk}
\end{figure}

The FDA Guideline \emph{Premarket Submissions for Management of Cybersecurity in Medical Devices}\footnote{https://www.fda.gov/regulatory-information/\\search-fda-guidance-documents/cybersecurity-medical-devices-quality-\\system-considerations-and-content-premarket-submissions} defines a security strategy based on the NIST Cybersecurity Framework, with the strategy of identify, protect, detect, respond and recover~(\cite{food2023cybersecurity, scherb2019execution}). A documentation about a hazard analysis and a trace matrix between cybersecurity controls risk is expected.
The risk should be managed and controlled as shown in figure \ref{fig:fdarisk}.
Following a market release, the FDA has the \emph{Postmarket Management of Cybersecurity in Medical Devices}, which describes postmarket measures, since cybersecurity is constantly evolving and it is not enough to only consider premarket. Postmarket measures can contain requirements for patching and updates. 

Beside theses two standards, there are further standardization defining risk and quality management but now further details about cybersecurity. 

ISO 13485\footnote{\url{https://www.iso.org/standard/59752.html}} is a quality management system specifically designed for for medical devices~(\cite{iso2016iso}). It defines the expectations towards quality management, management responsibility, resource management etc.

IEC 62304\footnote{\url{https://www.iso.org/standard/38421.html}} defines software life cycle processes as well as three different classes of medical devices~(\cite{jordan2006standard}). Class A defines medical devices where no injuries or harm to the patient are possible, even if the devices malfunctions. Class B defines medical devices, where non-serious injuries are possible, if the devices malfunctions and Class C defines devices where a malfunction can cause serious harm or death of a patient. 
The life cycle process standardizes the development process, the maintenance process, the risk management process, the configuration management process and the problem resolution process.

ISO 14971\footnote{\url{https://www.iso.org/standard/72704.html}}, a standard for application of risk management to medical devices to medical devices defines core components a medical devices manufacturer should consider to due identifying hazards and hazardous situations associated with a medical device as well as the evaluation, controlling and monitoring of appropriate risk management measures~(\cite{teferra2017iso}). 

\paragraph{Summary:} While we identified many guidelines that define how to assess and control cybersecurity risks, guidelines explaining how to design and test software for cybersecurity issues to reduce the risk in first place are rare, especially guidelines containing methods on how to securely develop software for CMDs are missing entirely. 

\subsection{Methods for Testing Cybersecurity}

For filling the gap in methods to check CMDs for cybersecurity, we will present existing background information how to detect vulnerabilities during the development and during the testing process. This includes methods to detect known vulnerabilities in libraries used by the CMDs as well as to detect unknown vulnerabilities which may be introduced during the development process. 

\subsubsection{CVE Search}

\begin{figure}
    \centering
    \includegraphics[width=0.5\textwidth]{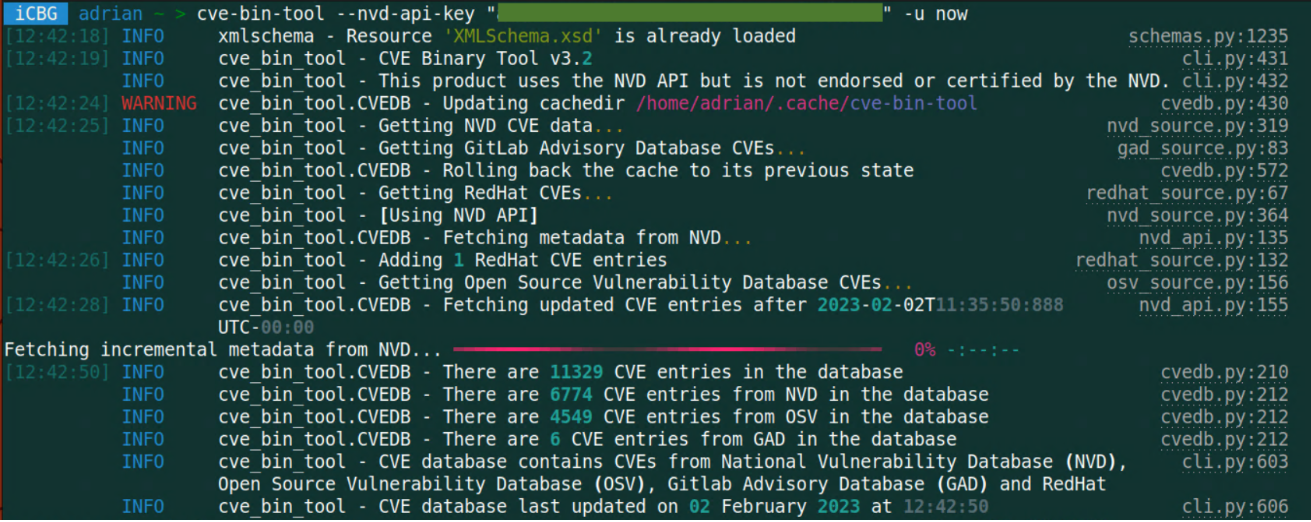}
    \caption{Output of the cve\_bin\_tool}
    \label{fig:cvebintool}
\end{figure}

CVE stands for Common Vulnerability and Exposure and is a standardized way to describe and to refer to known information security vulnerabilities and exposures. CVEs are collected in a database maintained by the Mitre Corporation, a non-profit organization. The aim of the CVE database is to give an overview over the weaknesses in certain software versions, such that companies know when it is essential to update. This is especially useful if a software product uses third party libraries of specific versions. The programmer can use the CVE database to check if any third party dependencies has new known vulnerabilities and decide when to update it. However, modern software has often hundreds of dependencies and it is already hard to check all dependencies in the first place, and even more complicated to monitor them permanently for weaknesses. Therefore, automatic search tools can take over the job that scan projects for software versions marked as vulnerable in the CVE database. The \emph{cve\_bin\_tool}\footnote{\url{https://github.com/intel/cve-bin-tool}} is such an automatic scanner. It can run against all Unix based firmware and software and produces a list of software libraries and programms which contain known vulnerabilities. Figure \ref{fig:cvebintool} shows an example output of the cve\_bin\_tool. 

\subsubsection{Firmware Analyzer}

\begin{figure}
    \centering
    \includegraphics[width=0.5\textwidth]{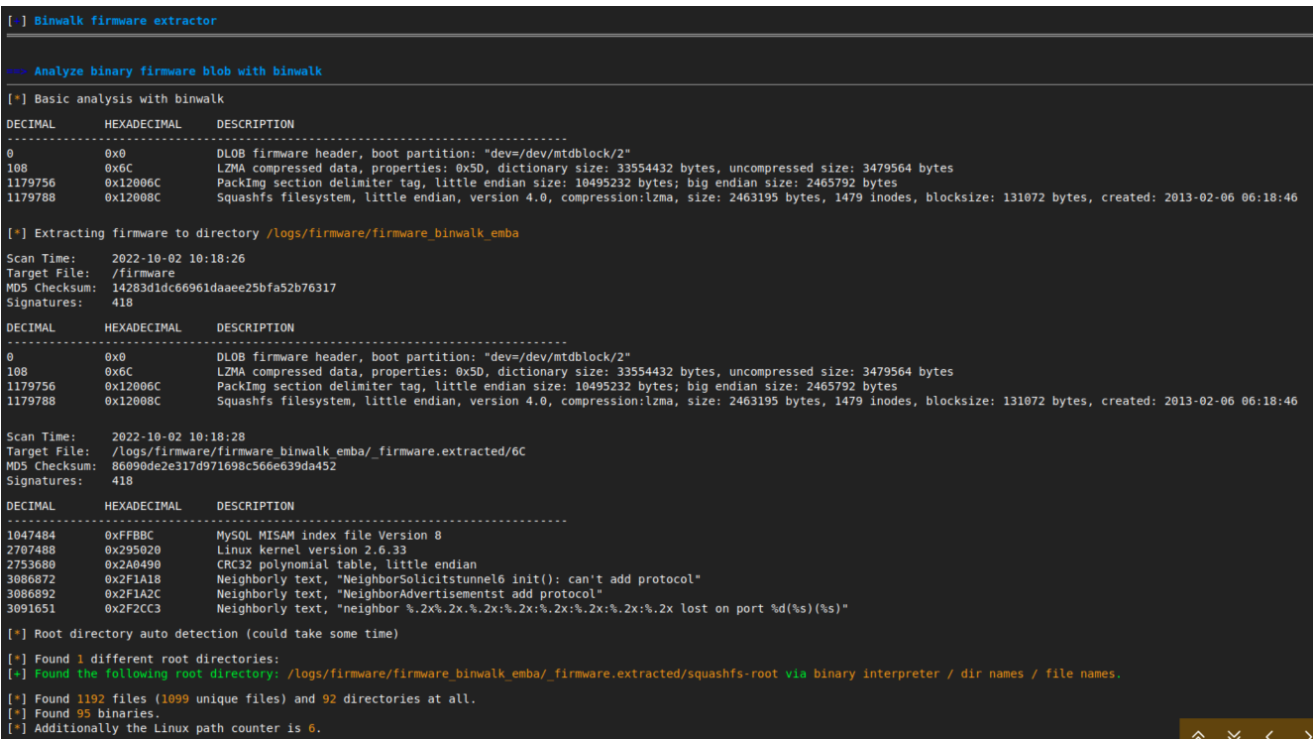}
    \caption{Output of EMBA}
    \label{fig:emba}
\end{figure}

Firmware analyzers can support developers of CMDs and other IoT devices to detect more vulnerabilities compared to the \emph{cve\_bin\_tool}~(\cite{zaddach2014avatar}). A firmware analyzer can find hard coded credentials, malware included in the firmware, the use of weak functions, weak permissions and even advanced methods like static analysis of included scripts~(\cite{schulz2018nexmon}). This wide range of analyzers enables the programmers to detect potential weaknesses and misconfigurations before shipping the product. Errors such as unchanged default passwords, which where set during the development process for convenience, can be detected and fixed, which is one of the most common errors. Moreover, many firmware analyzer can also extract and analyze the file system of ready-to-flash firmware. This way the built firmware as it would run on the device can be analyzed as well as enabling endusers different from the developers to analyze the firmware and get an impression of the security. The EMBA security analyzer for firmware of embedded devices is one of the most complete firmware analyzers for Unix based firmware systems. It contains many modules to detect various security flaws\footnote{\url{https://github.com/e-m-b-a/emba}} and generates sophisticated and detailed reports. Figure \ref{fig:emba} shows example output of EMBA. 

For non Unix based firmware, the analysis can be far more complicated. For statically linked firmware running directly on the hardware, e.g. insulin perfusors or pacemakers, with no or minimal operating system, reverse engineering is often the only way to understand which libraries are used, when no source code is available~(\cite{zaddach2013embedded}). In the case where the source code is available, as it would be for manufacturers and developers, \emph{cve-search}\footnote{https://github.com/cve-search/cve-search} can be used to automatically check the used library versions against the CVE database.

\subsubsection{Fuzzing}

Fuzzing is an efficient technology to detect unknown vulnerabilities in software, where generated inputs are passed to a computer program aiming to trigger crashes, vulnerabilities or other software flaws~(\cite{li2018fuzzing}). Fuzzing usually targets APIs and and all kind of interfaces where  data could be entered, since a cyber attack usually requires some kind of interaction or data flows. 
There are different variants and implementations of fuzzing but the concept is quite similar. Inputs are generated based on either the mutation of test inputs or the mutation of inputs generated by a grammar or structure. The inputs are passed to the computer program to be executed, and the behavior of the computer programs is monitored as well as the parts of the computer program triggered by the input.  
Based on the monitoring information, a fuzzer can decide if a generated input is worth keeping or if it is deleted. Usually a fuzzer keeps input which triggered previously unseen parts of the computer program, to maximize the code coverage and the change to find a flaw in the software~(\cite{fioraldi2020afl}). Beside the mutation based and generation based fuzzers~(\cite{miller2007analysis}), there are also hybrid fuzzers which try to combine constraint solving or concolic execution with fuzzing~(\cite{stephens2016driller, godefroid2012sage}). 

Popular fuzzers are \emph{AFL++}~(\cite{fioraldi2020afl}), which is one of the most used mutation based fuzzer as well as \emph{hongfuzz}\footnote{https://github.com/google/honggfuzz}. Since every fuzzing task is unique, fuzzers usually need to be adapted to the task given. For example, \emph{LibAFL} offers good customization options and makes it easier to adapt a fuzzer to a specific task~(\cite{fioraldi2022libafl}). Fuzzing can also applied directly on embedded devices using hardware in the loop testing techniques~(\cite{dunne2022powertrace, bacic2005hardware}).

\begin{figure}
    \centering
    \includegraphics[width=0.5\textwidth]{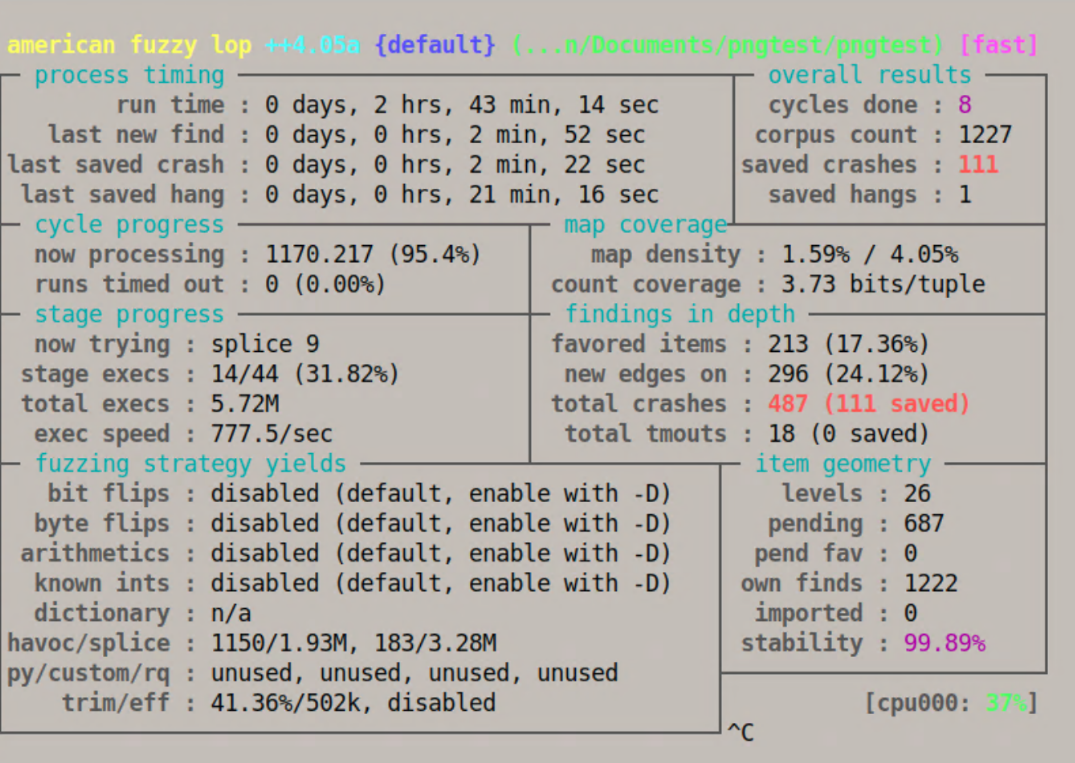}
    \caption{Fuzzing with AFL++}
    \label{fig:afl}
\end{figure}

Figure \ref{fig:afl} shows AFL++ in operation.

\subsubsection{Symbolic Execution}

Symbolic Execution is a formal method to analyze the behavior of computer programs by executing them symbolically, considering any non determined input to be a symbolic variable which could have any possible value~(\cite{king1976symbolic, baldoni2018survey}). In symbolic execution, all reachable paths are executed and non-reachable paths are ruled out by solving the path constraints. While being very slow for larger program due to the high number of different paths and combinations of paths a computer program could go, when feasible symbolic execution can prove the absence of certain bugs. For example, Symbolic Execution can detect paths, where the Instruction Pointer of a computer program depends on input instead of the program flow, which clearly is a vulnerability. Moreover, Symbolic Execution can generate test cases which test all possible outcomes using the constraint solver. Popular tool-kits for symbolic execution are Klee~(\cite{cadar2008klee}) and angr~(\cite{shoshitaishvili2016state}). 

\subsubsection{Weak Functions}

Weak functions, also called vulnerable functions, are functions within standard libraries which either cannot be used safely or are likely to be used in an unsafe manner~(\cite{almansoori2020secure}). Therefore, the best option for cybersecurity is to avoid these functions. These weak functions are widely existing in the C/C++ standard libraries~(\cite{yang2018source}). Since C/C++ are very common languages for developing embedded devices such as CMDs, vulnerabilities introduced by weak functions are a real problem. While there are more secure programming languages such as Rust and Ada~(\cite{barnes1995ada, boebert1985secure}), C/C++ are the most widespread. 
In the C/C++ standard library there are functions such as \textit{gets}, which cannot be safely used and there are functions like \textit{strcpy} or \textit{sprintf} which are highly dangerous if used wrong. Even functions like \textit{strncpy} or \textit{snprintf} which are considered to be secure, have flaws, and the programmer needs to make sure to use the functions as intended to avoid vulnerabilities. For example by not considering the string ending \verb|\0| at the end vulnerabilities can be introduced. 

\section{Methodology}
\label{sec:methodology}
The primary method we adopted is the Design Science Research (DSR) as proposed by Hevner and Chatterjee~(\cite{hevner2010design}), which is designed to create frameworks in information systems, which perfectly fits to our research question. DSR consists of multiple iterations to collect knowledge and develop a prototypical solution. 
In a first iteration we where collecting methods for checking cybersecurity of software in general as well as existing standards and guidelines for cybersecurity of medical devices. The goal of these phase is to identify the relevant methods for testing cybersecurity of software in general. We were searching on Google Scholar, Web of science, Science Direct IEEE and ACM with the keywords: "cybersecurity testing for medical devices", "guidelines for cybersecurity of medical devices", "testing for cybersecurity", "testing for vulnerabilities", "frameworks for cybersecurity of medical devices".
In our related work section we provide a literature review over the existing standards and guidelines. 

During our research, the identified methods for testing  were tried against the special requirements of CMDs. Therefore, we were focusing on how complicated the setup for the methods is as well as how well established it is for security testing. Next we ran own tests to see, if we can find vulnerabilities in CMDs using these technologies, using firmware and operating systems actually used in medical devices\footnote{\url{https://resources.sw.siemens.com/en-US/white-paper-using-linux-in-a-medical-device}}. We analyzed the results to find the most efficient tools to detect vulnerabilities and created a framework out of it. 
In a second iteration we went to domain experts to discuss our framework and further improve it. We had semi-structured interviews with ten different experts with the following professions: IT System Engineer, Research Assistant for Computer Science, Professor of Cybersecurity and Blockchain, Professor in Cybersecurity and Data privacy, Health IT Specialist, Head of Security Operations, Co-founder of a Cybersecurity Startup and a medical doctor. Based on their recommendation we created our CyMed-framework. 

\section{Framework for Testing CMDs}
\label{sec:framework}
\begin{figure}
    \centering
    \includegraphics[width=0.31\textwidth]{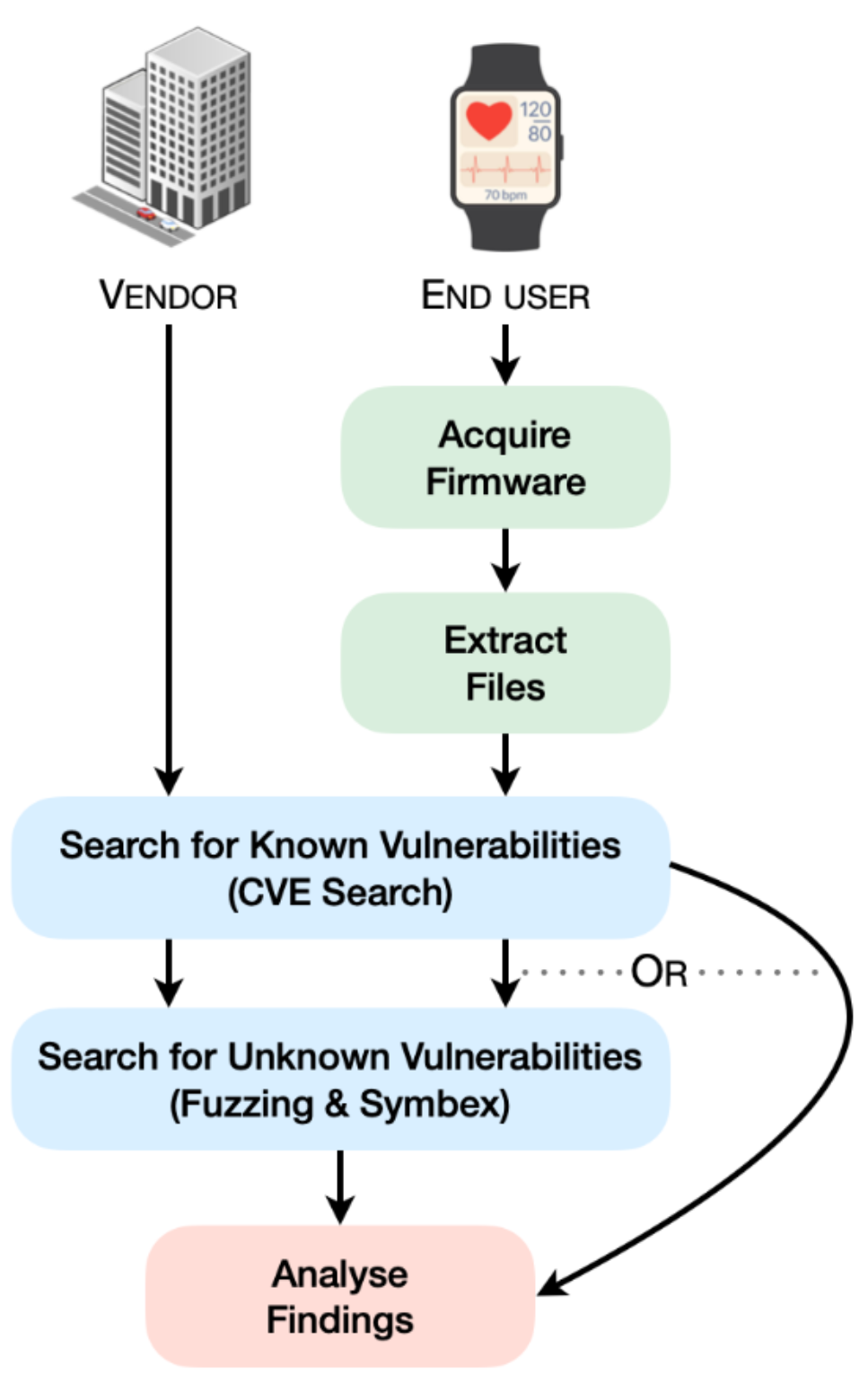}
    \caption{The CyMed-framework to strengthen cybersecurity of CMDs}
    \label{fig:framework}
\end{figure}

In this section we will describe our framework called CyMed to improve the resistance of CMDs against cyber attacks and modern cyber threats. 
CyMed focuses on manufactures and developers of CMDs but not exclusive, it also explains operators and end users how cybersecurity tests can be performed, even in absence of available source code. 
The focus of CyMed is rather preventing cyber attacks against the devices than data protection of patients data, but of course, preventing cyber attacks will also strengthen data protection and data integrity. 
Figure ~\ref{fig:framework} gives an overview over the CyMed-framework. 

The CyMed-framework has two different start points, depending if you have access to the source code of the CMD or not. Without source code availability, an additional step is required to obtain the (binary) firmware of the CMD and to extract the components to be tested from the firmware.
For a vendor or an CMD where the firmware is open source available, this step is not required. The next step of CyMed is to search for known vulnerabilities in the CVE database which may have been introduced by the use of third party libraries. Furthermore, in this step the configuration of the CMD should be analyzed and it should be made sure, that the devices is properly configured, no artifacts from the development process such as default passwords are left on the device. Afterwards, it is time to search for previously unknown vulnerabilities in the code as well as in the third party libraries of the CMD. The results of both the CVE Search and the search for unknown vulnerabilities will give an overview over potential issues in the code of the CMD and can give the manufacturer valuable information on how the security of the CMDs can be improved. Concrete decisions about how bugs and vulnerabilities are addressed and fixed are done in a last phase. In the following we will give more details about each phase. 

\subsection{Acquiring Firmware}
The analysis of a firmware requires access to the latter. For vendors this is generally not a problem since they have direct access to the firmware and the source code. Vendors are the main target group of CyMed, however, not exclusively. Operators may also want to assess the security of CMDs before adding them to their network. To assess the cybersecurity it is essential for an operator to acquire the firmware of the device. The easiest way may be to have a good relation with the manufacturer and get the firmware directly from them.
However, in most cases it is not so easy, therefore the firmware needs to be either acquired from update sites, from exploiting the update mechanism of the device or from extracting it from the devices. These tasks are unlikely to be practical for non cybersecurity trained professionals, limiting the possibilities to assess the security of the firmware to specialists. During our experiments, we found many vendors of CMDs are actively trying to keep the firmware from being analyzed, by for example not allowing operators to update the firmware themselves but requiring the devices to be send to the vendor for updates. 

\subsection{Extracting the Firmware}
\label{susec:extractfirmware}
If a firmware file could be obtained, the firmware needs to be extracted. Here are two cases. Many firmware are built on top of (real time) Linux or Unix systems. A Linux or Unix firmware can be extracted using tools like \emph{binwalk} or \emph{firmwalker}. \emph{binwalk} analyzes a dumped firmware file and can extract the Unix file system. 

\subsection{Search for Known Vulnerabilities}
With the firmware extracted the next step is to check if any known vulnerabilities are present in it. To do so all executable files and libraries contained in the firmware is checked against a Common Vulnerabilities and Exploits (CVE) database. There are several tools freely available that check software against known vulnerabilities. If \emph{binwalk} has been used in the previous step \ref{susec:extractfirmware}, which only extracts the firmware, \emph{cve-bin-tool} can be used to determine whether known vulnerabilities are present. This tool accesses the National Vulnerability Database API to check the files and generates a CVE dashboard that lists any findings of vulnerabilities. Alternatively the tool \emph{EMBA} can be used which extracts the firmware and then also directly analyzes it. Compared to \emph{cve-bin-tool} it creates a more comprehensive report but since it also performs more analysis it needs more time to fully analyze the firmware. 

\subsection{Search for Unknown Vulnerabilities}
While this step is not necessarily mandatory it has the potential to drastically increase the security and reliability of a CMD. Especially vendors should take the time and effort to further test their software to find unknown vulnerabilities hidden in their software. There exist many analysis techniques that are capable of analyzing software but for CyMed we propose the dynamic techniques fuzzing and symbolic execution. Dynamic analysis means that the targeted software runs during the analysis which usually gives better results than than static analysis, which only considers the source code. In the following we will further describe how these techniques can be used within CyMed by both vendor and end user.

\subsubsection{Fuzzing}

Since its invention fuzzing has been a successful technique to efficiently find vulnerabilities. But it is important to note that it can not guarantee the absence of them. As a vendor with access to the source code the code can be compiled with additional instrumentation which allows fuzzers such as \emph{AFL++} to track how much of the code was covered by the generated test cases~(\cite{fioraldi2020afl}). This allows for more efficient testing and can be used during development of the software to further harden it before it is released to CMDs. The same can be done as end user even if there is no access to the source code and it is possible to adapt \emph{AFL++} easily for firmware fuzzing~(\cite{zheng2019firm}). 
Fuzzing is an activity with an undefined ending. Fuzzing usually cannot achieve 100\% coverage, of the computer program, thus we always have a trade off between time and efficiency. For manufacturers it is recommended to apply the fuzzing process for a certain amount of time, for example when new code is added to the project. For fuzzing checks during the development process targets need to be created similar to unit tests. If the fuzzing process finds a bug in the defined time span, the code is refused and the developer needs to find and fix the bug. Only if the code runs stable over the testing period it will be added to the project, similar as done with unit tests. 
Generally, the most important fuzzing targets are API functions, especially those used for network communication, since they are most exposed for attackers.

For operators and pentester the fuzzing activity can run for an indefinite time, till either bugs are found or till the tests are aborted.

\subsubsection{Symbolic Execution}

Similar as fuzzing, symbolic execution is an activity which can be run on code which is added to the project. While fuzzing is a non-deterministic open ending activity, symbolic execution systematically executes all parts of a computer program and thus is deterministic. Symbolic execution is generally slow, but the beauty is, that it can be used as a formal method to prove the absence of certain bugs. Because of the slow execution it does not make sense to add symbolic execution to the build pipeline, but rather use it at the end of development process before shipping the CMD to verify one last time that the software is robust against cyber attacks. Setting up symbolic execution can happen in a similar way as fuzzing, as the tester also wants to start with API functions and trace the control flow from there to find which parts of the computer program or firmware are reachable by API calls and if on these paths there are any vulnerabilities. 

\subsection{Analyze findings}

At the very end of the process of checking CMDs, the findings need to be analyzed and the right actions need to be taken. For the search of known vulnerabilities, the results can usually be resolved with an update by the vendor or by following the recommendations of vendors or regulatory bodies to fix the updated security issues or even by the vendor themselves if the vulnerability is not in a third party library. In case there is not a patch against a known vulnerability available or it cannot be produced within reasonable time, the risk of the situation needs to be assessed. If the devices are connected directly to the internet, it may be required to take them offline until a patch is available. In other cases it may be enough to hide the devices in segments independent of rest of the devices on the operator's network. 

\begin{figure}
    \centering
    \includegraphics[width=0.5\textwidth]{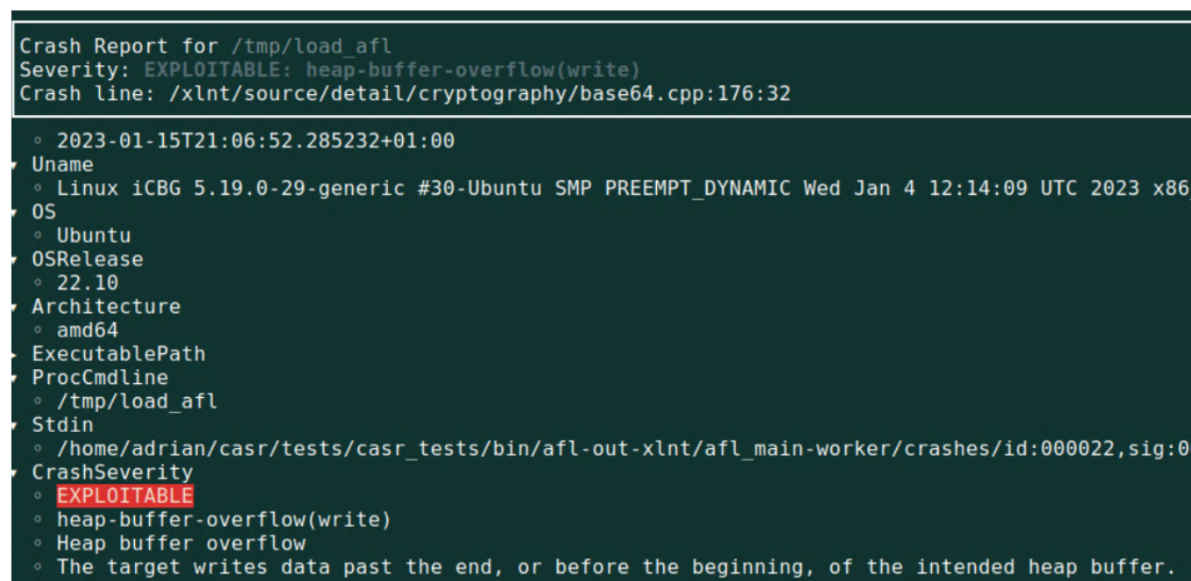}
    \caption{A triage tool (here: CASR) to find the security relevant crashes}
    \label{fig:casr}
\end{figure}

For previously unknown bugs and vulnerabilities the vendor needs to investigate them themselves. However, there are tools to support them. Before jumping to the debugger and analyzing every crash found by a fuzzer or by symbolic execution manually, a technique called triage could be used~(\cite{savidov2021casr}). Thereby, triage tools automatically analyses found crashes and check them for exploitability and impact. CASR (see figure \ref{fig:casr}) is an very efficient tool for triage of fuzzing results by estimating the severity and the exploitability.  
Bugs which can be exploited should be prioritized to ensure the security of the device. But also bugs which have a large impact and thus endanger the safe usage of the device need to be prioritized, depending on how likely a safety impact is. 

\section{Evaluation}
\label{sec:eval}

Our evaluation consists of two parts. In a first phase we analyzed the capabilities of the tools we identified in a previous literature review to find bugs in real-world firmware and operating systems designed for medical devices. In the second phase we evaluate the CyMed-framework by experts interviews.

For the first phase, we started with the phase to acquire firmware from IoT devices and CMDs for our experiments. These experiments are not a benchmark for how good different tools are working but just a validation of their general usability.
The tools we choose for CyMed are based on a literature review and have proven themselves to be efficient to find bugs and vulnerabilities in software. Our conclusion from these tests is, that \emph{EMBA} and fuzzing are strong tools which we would recommend everyone developing software for CMDs to try.
One of the focuses when testing the tools was if it is reasonable for software engineers to add their CI-tool chain. From our experience and from the literature review perspective, this question can be answered with yes. Our experiments showed us, that it is possible to setup fuzzing with a reasonable additional effort, while running tools such as \emph{EMBA} or \emph{cve-bin-tool} is even for less technical educated workers easily possible. 

The second and more important part of our evaluation where semi-structured expert interviews, which where mainly conducted online. The interviews started with explaining the CyMed-framework for testing cybersecurity of CMDs. Later, the experts where ask if they can follow the framework and how relevant they believed each step to be useful for testing for security. All respondents understood every step in the CyMed-framework based on the schematic and a short description provided to them. During the interviews two expert noted, that the steps are very complex to be performed for operators or other stakeholder without access to the source code. But at the same time they acknowledged that it is a good and functional strategy for vendors to test without source code. Though for testing without source code, cybersecurity experts would be required, but this is always the case if the access to the source code is not possible. One expert noted that applying such a framework to CMDs could be an advantage for vendors when used in advertisement.
A cybersecurity professor mentioned, that fuzzing is a very effective and successful technology in finding vulnerabilities and especially combined with an search for already known vulnerabilities with tools like \emph{cve-bin-tool}, however, fuzzing, while being a strong testing method,is not the only tool to be used. For example, manual reviews of important parts such as cryptography-implementations should not be neglected over fuzzing. To ensure the correctness of network and cryptographic protocols formal methods such as model checking may be required to ensure correctness~(\cite{li2013offline}; \cite{cicotti2015towards}). Since model checking is very costly in creating and executing the models, it is only realistic to be used in very critical subsets. Moreover, the professor pointed out, that it is important to also focus on normal unit testing in a way, that not only functionality is tested but also the failing of function is handled correctly, since this is a common source of errors. 
A Swiss medical doctor mentioned during the interview, that they do not receive any training how to secure the devices and that they have to trust devices they get provided as is. Meanwhile the health care sector gets more and more digital with the introduction of electronic patient records in Switzerland, the EU and other countries. 
All experts agreed the CyMed-framework highlights the most important state-of-the art technology to test for cybersecurity and that applying the framework to CMDs will not only increase their security but also their reliability. They also mentioned that CyMed would require regular updates when the technology for testing software evolves.

The responses are aligned with our expectations and prove the relevance of our cybersecurity framework CyMed to provide guidelines for better and more sustainable development of CMDs.

\section{Conclusion \& Future Work}

In this paper we present and evaluate a comprehensive framework -- CyMed -- for increasing the cybersecurity of CMDs. While most frameworks refer to state-of-the-art security and do not contain concrete measure and steps a vendor could follow to increase their cybersecurity, CyMed aims to fill this gap. To build the CyMed-framework we started with an literature review to identify the required tools and related resources as well as to point out our research gap. We tried all tools ourselves to verify their ability to find bugs and vulnerabilities. Based on our experiences we created a framework with which a developer of CMDs could test their devices. Afterwards, the framework was evaluated by experts interviews. The results of the interviews show that CyMed is addressing a need by the industry as well as the relevance of the chosen tools for testing for cybersecurity of CMDs.
The CyMed framework is the answer to our first research question (section \ref{sec:ProblemStatement}). It contains also the answers to the second research question by giving concrete suggestions and tools to improve the resistance of CMDs against cyber attacks. 

Beside increasing the security of CMDs themselves using the discussed mechanisms, it is also required to secure the hospital or the organization using them. Therefore, different methods such as the NIST framework can be used~(\cite{nistframework}, \cite{Society5.02023:Agile_Management_in_Cybersecurity, scherb2018smart}).  Moreover, comprehensive data protection, transport and sharing of medical data is required~(\cite{zhang2016sharing, scherb2017network}, \cite{10.1145/2984356.2984366, scherb2017execution}). 
Also it make sense to use modern technologies to educate the operators and patients in securely using CMDs and the entire IT-infrastructure around it. Therefore different forms of awareness campaigns can be used, for example phishing campaigns, seminars or serious games~(\cite{scherb2023cyber, scherb2023serious, sifalakis2014information}). 
The integration of such methods into CyMed to provide a more holistic approach to the security of CMDs is the logical next step. However, for now, our framework focuses on the cybersecurity of the software running on CMDs or of software which is a CMD itself. 
In future, CyMed needs to be kept up to date with the development of security testing tools.

\bibliographystyle{unsrt}  
\bibliography{main}

\end{document}